\def\ARXIVVERSION{1}
\documentclass[letterpaper]{article} 
\usepackage{aaai25}  
\usepackage{times}  
\usepackage{helvet}  
\usepackage{courier}  
\usepackage[hyphens]{url}  
\usepackage{graphicx} 
\usepackage{natbib}  
\usepackage{caption} 
\usepackage{algorithm}
\usepackage{algorithmic}

\usepackage{newfloat}
\usepackage{listings}
\DeclareCaptionStyle{ruled}{labelfont=normalfont,labelsep=colon,strut=off} 
\floatstyle{ruled}
\newfloat{listing}{tb}{lst}{}
\floatname{listing}{Listing}
\usepackage{booktabs}
\usepackage{multirow}
\usepackage{amsmath,amssymb}

\title{TCAM-Diff: Triplane-Aware Cross-Attention Medical Diffusion Model}
\author{
    Zhenkai Zhang,
    Krista A. Ehinger,
    Tom Drummond
}
\affiliations{
    School of Computing and Information Systems, The University of Melbourne\\

    zhenkai.zhang@student.unimelb.edu.au, \{kris.ehinger, tom.drummond\}@unimelb.edu.au
}

\ifdefined\ARXIVVERSION
\nocopyright
\newcommand{\arxivstatusnote}{%
  \begingroup
  \renewcommand{\thefootnote}{}
  \footnotetext{Accepted at the Thirty-Ninth AAAI Conference on Artificial Intelligence (AAAI 2025).}
  \addtocounter{footnote}{-1}
  \endgroup}
\else
\newcommand{\arxivstatusnote}{}
\fi

\begin{document}

\maketitle
\arxivstatusnote

\begin{abstract}
We introduce TCAM-Diff, a novel 3D medical image generation model that reduces the memory requirements to encode and generate high-resolution 3D data. This model utilizes a decoder-only autoencoder method to learn triplane representation from dense volume and leverages generalization operations to prevent overfitting. Subsequently, it uses a triplane-aware cross-attention diffusion model to learn and integrate these features effectively. Furthermore, the features generated by the diffusion model can be rapidly transformed into 3D volumes using a pre-trained decoder module. Our experiments on three different scales of medical datasets, BrainTumour $128\times128\times128$, Pancreas $256\times256\times256$, and Colon $512\times512\times512$, demonstrate outstanding results. We utilized MSE and SSIM to assess reconstruction quality and leveraged the Wasserstein Generative Adversarial Network (W-GAN) critic to assess generative quality. Comparisons with existing approaches show that our method gives better reconstruction and generation results than other encoder-decoder methods with similar sized latent spaces.
\end{abstract}

\begin{figure*}[t]
\centering
\includegraphics[width=1.0\textwidth]{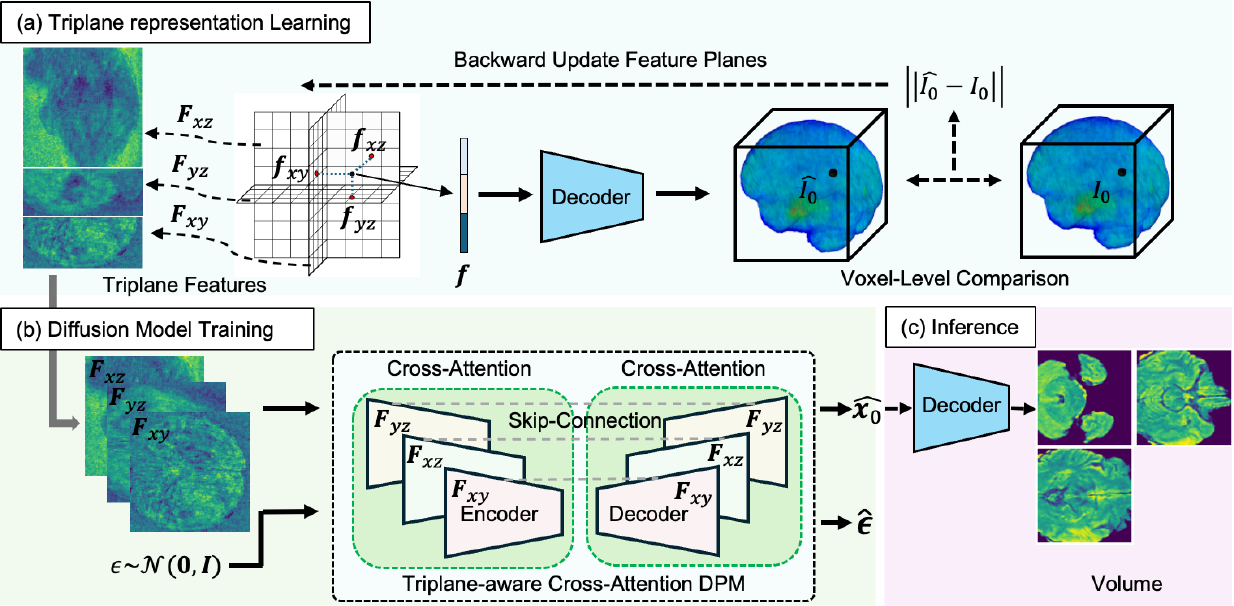} 
\caption{The overview of our two-stage model architecture. (a). Triplane representation learning: triplane features \{$\mathbf{F}_{xy}$, $\mathbf{F}_{yz}$, $\mathbf{F}_{xz}$\} are learned using a decoder-only method with voxel-level comparison. (b). Diffusion model training: triplane features are processed with triplane-aware cross-attention and skip connections in the diffusion model. (c). Inference: the diffusion model's outputs are decoded by the pre-trained decoder to reconstruct 3D volumes. }
\label{fig1}
\end{figure*}

\section{Introduction}
The development of generative models has revolutionized numerous fields, ranging from natural language processing to image synthesis. These models not only enhance data diversity by generating synthetic examples of rare classes, improving classifier robustness, but also facilitate conditional generation. This allows for the creation of data constrained to match specific measurements, offering plausible interpretations and expanding the analytical capabilities across various applications. 

Recently, diffusion models~\cite{ho2020denoising} have gained attention for their ability to generate high-quality realistic data by effectively modeling complex distributions. Their flexibility and robustness make them suitable for a wide range of applications. Building on this, latent diffusion models (LDM)~\cite{rombach2022high} extend these capabilities by operating in a lower-dimensional latent space, which reduces computational complexity and broadens their applicability to more diverse and complex tasks, such as image superresolution~\cite{zhao2023partdiff}, text-to-image generation~\cite{rombach2022high}, 3D object generation~\cite{shue20233d,ntavelis2023autodecoding}, medical image synthesis~\cite{kwon2019generation,Pinaya2022BrainIG,Khader2022MedicalDD}, video generation~\cite{wang2023lavie}, and speech synthesis~\cite{liu2023diffvoice}.

However, despite substantial progress, existing generative techniques~\cite{segato2020data,chong2021synthesis,Pinaya2022BrainIG,Khader2022MedicalDD} often fail to deal with 3D medical data, such as CT and MRI, particularly when it comes to maintaining high resolution and anatomical accuracy due to the prohibitive memory footprint these entail. \cite{kwon2019generation,Pinaya2022BrainIG,Khader2022MedicalDD} propose methods that utilize autoencoders to compress 3D medical volumes into a latent space, followed by training with latent generative models to reduce the memory footprint required for processing such data. However, due to the inherent limitations of the autoencoder architecture, including loss of critical detail during compression and the dual-step encoding-decoding process, these models still face challenges when dealing with high-resolution medical data. 

To address this issue, we propose a novel approach that utilizes a latent diffusion generative model operating on a triplane representation~\cite{chan2022efficient} to learn complex structures within 3D medical data effectively. Additionally, in the process of learning triplane representations, we utilize the decoder-only architecture, which significantly reduces memory usage, allowing the model to handle higher-resolution medical data more efficiently. This is particularly advantageous because the diffusion model focuses solely on the decoding process. Furthermore, while previous applications of triplane representations have been limited to modeling the surfaces of 3D objects~\cite{chan2022efficient,shue20233d}, our work demonstrates that this approach can also effectively represent dense volumes, such as medical datasets.

The main contributions of this paper can be summarized.
\begin{itemize}
    \item We introduce a \textbf{decoder-only} \textbf{autoencoder} that decodes latent triplane representations \cite{chan2022efficient} into dense 3D volumes using iterative backpropagation. Our approach outperforms baseline models like VAE-GAN \cite{Pinaya2022BrainIG} and VQ-GAN \cite{Khader2022MedicalDD} in reconstruction quality, with equivalent latent dimensions. Additionally, our model reduces memory usage during training, allowing for the processing of higher resolution images compared to baseline models.
    
    \item Our method introduces a novel diffusion model named TCAM-Diff ({\bf T}riplane-aware {\bf C}ross-{\bf A}ttention {\bf M}edical {\bf Diff}usion model) that uses convolutional layers to operate on each of the three orthogonal feature planes independently and specially designed transformer layers to integrate shared information across them. 
    
    \item To objectively compare our method with others, we present an approach that uses a {\bf W-GAN critic}~\cite{Arjovsky2017WassersteinG} for evaluating the relative performance of generative models in novel domains. This approach estimates how closely the models' samples mimic real data, offering a robust metric for assessing generative quality.
\end{itemize}

\section{Related Work}
Diffusion models have achieved state-of-the-art results in non-medical fields, showcasing their advanced capabilities in image generation and data synthesis across various applications~\cite{yang2023diffusion,phung2023wavelet}, especially, the advent of latent diffusion models~\cite{{rombach2022high}} has furthered this progress by optimizing within the latent space to increase the resolution of generated images. This success has sparked efforts to apply these models within the medical field, aimed at enhancing the diversity of medical data~\cite{Khader2022MedicalDD,qin2023class,Pinaya2022BrainIG}, facilitating the conditional generation of medical data~\cite{qin2023class}, and improving the performance of downstream medical tasks~\cite{amirhossein2022diffusion}.

Triplane representation~\cite{chan2022efficient} is a method that encodes 3D data by projecting it onto three orthogonal planes, effectively reducing the complexity of 3D structures into manageable 2D representations. This approach captures the essential features of a 3D object by learning from its projections on the $\{XY, YZ, ZX\}$ planes, enabling efficient spatial information modelling. The benefits of triplane representation include reduced computational complexity, as it simplifies 3D data processing into lower-dimensional spaces, and the ability to preserve essential structural information while minimizing the memory footprint. Consequently, triplane representations~\cite{chan2022efficient} have been widely employed in facial 3D reconstruction~\cite{chan2022efficient} and 3D object modelling~\cite{shue20233d}, primarily used to capture surface information. However, research on applying triplane representation to dense volumes, such as medical data, is noticeably lacking. This gap likely exists because dense volumes contain significantly more complexity and detailed features compared to the surface representations that triplane-related methods typically address. Our work demonstrates that triplane representations can also effectively represent dense volumes.

Medical image data, such as CT and MRI, presents greater challenges compared to two-dimensional image data, due to its volumetric attributes. The intuitive approaches are modelling the medical data directly with GAN~\cite{kwon2019generation,hong20213d,ozbey2020three,cirillo2021vox2vox}, or diffusion models~\cite{dorjsembe2022three}. However, these generative models in the medical field are often limited by memory constraints, and struggle to handle large-scale or high-resolution medical datasets. Recently, several advanced models~\cite{Pinaya2022BrainIG,Khader2022MedicalDD} have worked on utilizing the 3D autoencoder models to extract the latent dimensions from medical data, as the input of the 3D latent diffusion models, to generate the high-resolution images. Despite the advancements, such models still suffer the limitations of 3D autoencoder models, such as high computational load, which takes longer training times and higher demands on hardware, memory intensity, complexity in training, etc. In contrast, our approach utilizes a decoder-only architecture which effectively reduces memory usage, and computational load during the training and inference. This method not only simplifies the model structure but also enhances efficiency, particularly when handling large-scale or complex datasets.

\section{Methodology}
Our model, shown in Figure~\ref{fig1}, consists of two stages. First, a decoder-only autoencoder compresses 3D medical images into a triplane representation~\cite{chan2022efficient} using iterative backpropagation for encoding. This approach is advantageous because, for generative models, decoding speed is critical, while slower encoding is acceptable during training. In the second stage, a triplane-aware cross-attention 2D diffusion model learns from the triplane features, effectively integrating them into the diffusion process.

\subsection{Triplane Representation Learning}
The triplane representation optimally integrates implicit and explicit modelling advantages by maintaining three 2D feature planes in memory, rather than an entire voxel grid. This configuration enhances memory efficiency and accelerates the prediction process~\cite{chan2022efficient}. Here, our model uses the three orthogonal 2D feature planes $\mathbf{F}_{xy, \theta_{xy}} \in \mathbb{R}^{C \times H \times W}$, $\mathbf{F}_{yz, \theta_{yz}} \in \mathbb{R}^{C \times W \times D}$, $\mathbf{F}_{xz, \theta_{xz}} \in \mathbb{R}^{C \times H \times D}$  to represent a 3D dense medical volume, where the $(H,W,D)$ is the shape size of the 3D medical volume, $C$ is the feature channels for each feature plane, and $\theta_{xy}, \theta_{yz}, \theta_{xz}$ mean these three planes are parametric planes that are learned during the training process, and a compact decoder module, typically an MLPs (Multilayer Perceptron), is used to convert features sampled from these planes into corresponding spatial intensities.

We predefined multiple parametric triplane representations, 
$$\left\{
\left(\mathbf{F}_{xy}^{(0)},\mathbf{F}_{yz}^{(0)}, \mathbf{F}_{xz}^{(0)}\right),\cdots,\left(\mathbf{F}_{xy}^{(i)},\mathbf{F}_{yz}^{(i)}, \mathbf{F}_{xz}^{(i)}\right),\cdots 
\right\}$$
based on the number of objects in the training set. Given a three-dimensional point $\mathbf{x}^{(i)}=(x, y, z)$, this point is projected onto each axis-aligned plane of the $i$-th predefined triplane, the corresponding feature values are queried as $\mathbf{f}_{xy}^{(i)}=\mathbf{F}^{(i)}_{xy}(x,y),\; \mathbf{f}_{yz}^{(i)}=\mathbf{F}^{(i)}_{yz}(y,z),\; \mathbf{f}_{xz}^{(i)}=\mathbf{F}^{(i)}_{xz}(x,z)$ and concatenated as $\mathbf{f}(\mathbf{x}^{(i)})=\text{concat}\left(\mathbf{f}_{xy}^{(i)},\mathbf{f}_{yz}^{(i)},\mathbf{f}_{xz}^{(i)}\right)$, and then decoded through an MLP (multi-layer perceptron) network with parameters $\theta$, to estimate the intensity value $\mathbf{\widehat{I}}(\mathbf{x}^{(i)})$ at point $\mathbf{x}^{(i)}$ of the $i$-th object. 
\begin{equation}
\mathbf{\widehat{I}}(\mathbf{x}^{(i)}) = \mathbf{\text{MLP}}_\theta(\mathbf{f}(\mathbf{x}^{(i)}))
\label{e0}
\end{equation}

During the training process of the triplane decoder-only model, we optimize both the MLP network and the parameters of the triplane feature planes. 

In our experiments, all objects in the same dataset share the same MLP network. Since all objects within the same medical dataset contain similar types of data and features, sharing the same MLP network among these objects leverages this similarity to enhance learning and generalization. This commonality ensures that the network can efficiently learn a generalized model capable of handling variations across similar instances without overfitting specific features of individual objects, and ensure that the outputted triplane features are consistent and of high quality across the entire dataset. 

\subsection{Loss Functions and Regularization Approaches}
\label{m1}

The foundational training objective for our model is the SmoothL1Loss~\cite{girshick2015fast}. This loss function, akin to the Mean Squared Error (MSE) Loss but less sensitive to outliers, calculates the discrepancy between the estimated intensity value $\mathbf{\widehat{I}}(\mathbf{x}^{(i)})$, and ground truth intensity value $\mathbf{I}(\mathbf{x}^{(i)})$. The formulation of the loss function is expressed as follows:
\begin{equation}
    \mathcal{L}_{basic} = \sum^N_i\sum^M_j{Smooth_{L1}\left(\mathbf{\widehat{I}}\left(\mathbf{x}^{(i)}_j\right)- \mathbf{I}\left(\mathbf{x}^{(i)}_j\right)\right)}
\end{equation}
in which
\begin{equation}
Smooth_{L1}(x) = \left\{ 
\begin{array}{ll}
\frac{x^2}{2\beta} & if\; |x| < \beta \\
|x| - \frac{1}{2\beta} & \text{otherwise}
\end{array}
\right.
\end{equation}
where $N$ is the total object number, $M$ is the total points sampled from an object, $i$ represents the $i$-th object, $j$ indicates the point within a single object, and in our experiments, the hyper-parameter $\beta$ was set to 0.3. 

While this approach is effective in reducing the average error across pixel values, it often overlooks aspects critical to human perception, such as texture, contrast, and structural integrity. Through our experiments, we found that images optimized only for SmoothL1Loss may appear blurry or lack fine details and visual fidelity perceptible to the human eye. 

To address these shortcomings, we augment our loss function framework with Perceptual Loss~\cite{johnson2016perceptual}, which contributes significantly to enhancing the quality of the reconstructed images beyond mere pixel accuracy. In addition, it also can increase generalization by focusing on high-level features rather than low-level pixel details.
\begin{equation}
\mathcal{L}_\text{feat}(x, y) = \sum_{l=1}^{L} \frac{1}{N_l} \|\phi_l(x) - \phi_l(y)\|^2
\end{equation}
where $\phi_l(x),\phi_l(y)$ are the activations of the $l$-th layer of a pre-trained neural network (using the pre-trained 'squeeze' network in our experiments), $N_l$ is the number of elements in the $l$-th layer, and $L$ is the total number of layers used for the loss calculation.

In addition to employing loss functions aimed at enhancing the performance of image reconstruction, ~\cite{shue20233d} propose using the total variation (TV) regularization terms can simplify the data manifold by eliminating unnecessary high-frequency information from feature planes, aligning their distribution with natural images. 
\begin{equation}
\mathcal{L}_{TV} = \mathbf{TV}(\mathbf{F}_{xy}^{(i)}) + \mathbf{TV}(\mathbf{F}_{yz}^{(i)}) + \mathbf{TV}(\mathbf{F}_{xz}^{(i)})
\end{equation}

We employ root mean square normalization (RMSNorm) \cite{zhang2019root} to ensure each feature plane maintains unit energy during training, providing consistent scaling across planes. Compared to direct scaling to fixed ranges like $[0,1]$ or $[-1,1]$, RMSNorm \cite{zhang2019root} preserves relative feature dynamics without imposing rigid bounds, preventing bias from varying input magnitudes and promoting stable, uniform learning in the diffusion model.
\begin{equation}
\bar{a}_i = \frac{a_i}{\mathbf{RMS}(a)}, \; \mathbf{RMS}(a) = \sqrt{\frac{1}{n} \sum_{i=1}^n a_i^2}
\end{equation}

To improve MLP generalization, we propose MLP Noise Regularization. This involves adding controlled noise to features during training. Since features are normalized to maintain unit energy, noise is carefully managed using the signal-to-noise ratio (SNR).

Then, the normalized and generalized features become,
\begin{equation}
\mathbf{f}(\mathbf{x})' = \mathbf{\text{RMSNorm}}(\mathbf{F})(\mathbf{x}) + C * \mathbf{\epsilon}
\end{equation}
where $C=0.32$, $\mathbf{\epsilon} \sim \mathcal{N}(0, \mathbf{I})$ can achieve better generalization in our experiments.

Our loss functions for training the triplane decoder-only model are described as follows:
\begin{equation} 
\mathcal{L}_{total}= \mathcal{L}_{basic} + \lambda_1\mathcal{L}_\text{feat} + \lambda_2 \mathcal{L}_{TV} 
\end{equation}
in which, $\lambda_1=$$1\mathrm{e}{-2}$, $\lambda_2=$$1\mathrm{e}{-3}$ in our experiments.

\subsection{Triplane-aware Cross-Attention Diffusion Model}
\begin{figure}[t]
\centering
\includegraphics[width=1.0\columnwidth]{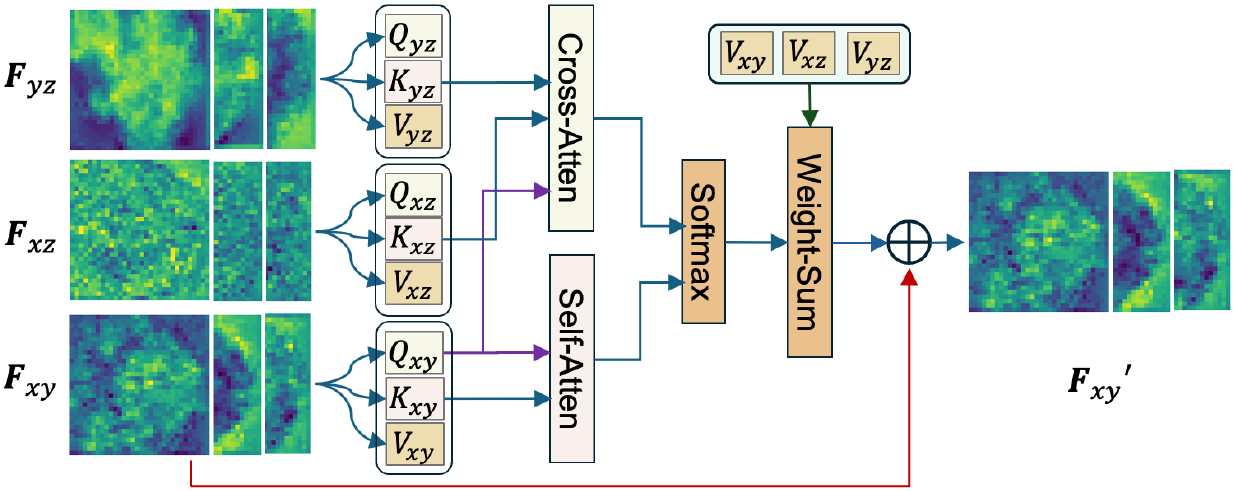}
\caption{The attention block for $\mathbf{F}_{xy}$, incorporating both self-attention and cross-attention mechanisms. The same structure is applied to the feature planes $\mathbf{F}_{yz}$ and $\mathbf{F}_{xz}$. The structure ensures the effective integration of information across different planes.}
\label{fig2}
\end{figure}
Upon acquiring meaningful triplane features, the subsequent phase involves leveraging a diffusion model to learn from these data representations.
The conventional method~\cite{shue20233d} concatenates three orthogonal 2D feature planes $(\mathbf{F}_{xy},\mathbf{F}_{yz}, \mathbf{F}_{xz})$ as the input of the diffusion model. While this concatenation technique simplifies data handling and reduces computational requirements, it introduces significant drawbacks. It does not preserve the intrinsic 3D spatial relationships inherent in the data, as the model processes the concatenated planes without recognizing their mutual dependencies. This limitation often leads to diminished accuracy in the reconstructed 3D structures and impairs the model's ability to comprehend the complexity of 3D spatial dynamics fully.

In Figure \ref{fig1} (b), we present a novel approach where each of the three mutually orthogonal feature planes is trained independently within the diffusion model. Although these planes are segmented, they originate from the same object and therefore share intrinsic information. To effectively leverage this shared data, we incorporate a cross-attention layer that facilitates bidirectional information flow between the planes. The structure of our newly designed attention blocks is depicted in Figure \ref{fig2}. This innovative strategy allows for the deep learning of each plane’s unique characteristics while simultaneously integrating knowledge across all planes, thereby significantly enhancing the model’s overall performance.

Our diffusion model is based on simultaneously estimating both the image and the noise~\cite{Zhang2023ImprovingDD}, the training objectives are:
\begin{equation}
    \min_\theta \mathbb{E}\left[ \|\mathbf{R}_\theta(\mathbf{x}_t, t) - \mathbf{x}_0 \| + \|\mathbf{\epsilon}_\theta(\mathbf{x}_t, t) - \mathbf{\epsilon} \| \right]
\end{equation}
where $\widehat{\mathbf{x}_0} = \mathbf{R}_\theta(\mathbf{x}_t, t)$, $\widehat{\mathbf{\epsilon}}=\mathbf{\epsilon}_\theta(\mathbf{x}_t, t)$, and $\mathbf{x}_0$ is the triplane representations with $\{\mathbf{F}_{xy}, \mathbf{F}_{yz}, \mathbf{F}_{xz}\}$.

\section{Experiments Setup}
\subsection{Dataset}
We conduct the experiments on three publicly available datasets with varying scales: the BraTS dataset includes 750 4D MRI volumes (484 training, 266 testing) with dimensions of $(240\times240\times155)$, covering four modalities: Fluid-Attenuated Inversion Recovery (FLAIR), T1-weighted (T1w), T1 with gadolinium contrast (T1gd), and T2-weighted (T2w)~\cite{menze2014,Simpson2019ALA}. The Pancreas Tumour dataset contains 420 3D CT volumes (282 training set, and 139 testing set) with varying dimensions~\cite{Simpson2019ALA}, The Colon Cancer dataset includes 190 3D CT volumes with 126 training and 64 testings, with diverse shapes of each scan~\cite{Simpson2019ALA}.

\subsection{Preprocessing}
Given that current 3D medical generative models~\cite{Pinaya2022BrainIG,Khader2022MedicalDD,liu2023inflating,friedrich2024wdm} are trained on single-channel datasets, we processed three datasets following VAE-GAN~\cite{Pinaya2022BrainIG} preprocessing methods to extract single-channel data for training. Using the MONAI toolkit~\cite{monai2020project}, we further preprocessed the datasets to validate the model's capabilities at different scales: the BraTS dataset was center-cropped to $128\times128\times128$, the Pancreas dataset was resized to $256\times256\times256$, and the Colon dataset to $512\times512\times512$. Intensity values were scaled to the $[0,1]$ range across all datasets~\cite{Simpson2019ALA}.

Additionally, utilizing current mainstream 3D medical segmentation models, which are trained on multi-channel data, to verify that our model's reconstructed results do not lose crucial information, we have also used the 4-channel multi-modal BraTS dataset~\cite{menze2014,Simpson2019ALA} as the fourth dataset in our experiments. This dataset contains three segmented labels: GD-enhancing Tumor (ET — Label 4), Peritumoral Edema (ED — Label 2), and Necrotic and Non-Enhancing Tumor Core (NCR/NET — Label 1)~\cite{menze2014multimodal,bakas2018identifying}.

\subsection{Baseline Models}
Our experiment utilized two baseline models based on architectures that integrate autoencoder with the latent diffusion model (LDM). One is a VAE-GAN combined with a 3D LDM~\cite{Rombach2021HighResolutionIS,Pinaya2022BrainIG}, which is reproduced by MONAI~\cite{monai2020project} based on the content of the paper, and the other is a VQ-GAN coupled with a 3D LDM (medical diffusion)~\cite{Khader2022MedicalDD}. We conduct comparative analyses on the quality and performance of autoencoder and generative models separately. Since the baseline model hasn't released its pre-trained models, the results were obtained by following the steps in the papers and retraining with the publicly available code.

\subsection{Evaluation Metrics}
Autoencoder reconstruction quality was evaluated using two metrics: MSE for pixel-level assessment and 3D SSIM \cite{Wang2004ImageQA} for structural-level evaluation. We also used SegResNet \cite{myronenko20193d} to compare the performance of our model's reconstructions with baseline models in downstream 3D medical segmentation tasks.

To verify the performance of the generative models, we utilized critic from W-GAN~\cite{Arjovsky2017WassersteinG}, to assess the distance between the real data distribution and the generated distribution. It measures how "far" the generated data is from the real data in terms of distribution, instead of just classifying whether data is real or fake in traditional GAN~\cite{goodfellow2020generative}. W-GAN critic~\cite{Arjovsky2017WassersteinG} assigns a score that represents the "earth mover’s distance" (also known as the Wasserstein distance) between the distribution of the generated data and the distribution of the real data. 
    $W(\mathbb{P}_r, \mathbb{P}_\theta) = \sup_{\|f\|_L \leq 1} \left( \mathbb{E}_{x \sim \mathbb{P}_r} [f(x)] - \mathbb{E}_{x \sim \mathbb{P}_\theta} [f(x)] \right)$
where $\mathbb{P}_r$ is real data distribution and $\mathbb{P}_\theta$ from generative model.
In practice, the lower Wasserstein distance means the closer the generated data's distribution to the real distribution, which indicates the model's ability to generate high-fidelity outputs. 

\cite{Pinaya2022BrainIG} used {\bf Fréchet Inception Distance (FID)}~\cite{heusel2017gans} to measure the performance of the generated model. However, they mentioned utilizing a pre-trained Med3D model~\cite{chen2019med3d} for feature extraction from 3D medical data but did not make the associated code publicly available. Moreover, there are issues with loading the provided pre-trained models, which will be questionable whether the extracted features are meaningful. Additionally, useful FID value must be computed in a large enough dataset. Due to the size of our training dataset, obtaining convincing results is challenging, leading us to decide against using FID as an evaluation metric.

\subsection{Implementation Details}
Our experiments were conducted on A100 GPUs, each with 80GB of GPU RAM. We trained the baseline models on 4 GPUs with default settings. Our decoder-only model was trained using only 1 GPU, and we utilized 4 GPUs to train the generative model. The hyperparameters in our experiments included an Adam optimizer with a $3\mathrm{e}{-5}$ learning rate for the BraTS dataset and a $1\mathrm{e}{-5}$ learning rate for higher-resolution datasets. The loss function weights were set to $\lambda_1=1\mathrm{e}{-2}$ and $\lambda_2=1\mathrm{e}{-3}$ for BraTS, and $\lambda_1=0$ for higher-resolution datasets. To ensure stability, we applied gradient clipping with a max norm of 1.0. Additional hyperparameters, architectures, and experimental details are provided in the appendix.

\section{Experimental Results}
In the following, VAE-GAN and VQ-GAN refer to the autoencoder process, while VAE-GAN-LDM~\cite{Pinaya2022BrainIG} and VQ-GAN-LDM~\cite{Khader2022MedicalDD} refer to the complete models, including both the autoencoder and generative components.

\subsection{Autoencoder Models}
\begin{table*}[h]
\centering
\begin{tabular}{lcccccc}
\toprule
Method & MSE $\downarrow$ & 3D SSIM $\uparrow$ & Latent Dimension & Parameters & GPUs & Time \\
\midrule
VAE-GAN & 0.0015 & 0.8720 & 2 $\times$ (8, 32, 32, 32) & 524,288 & 4 & 40h \\
VAE-GAN* & 0.0014 & 0.8818 & 2 $\times$ (8, 32, 32, 32) & 524,288 & 4 & 38.5h \\
VQ-GAN & 0.0079 & 0.8393 & (16, 32, 32, 32) & 524,288 & 4 & 37h \\
VQ-GAN* & 0.0045 & 0.8785 & (16, 32, 32, 32) & 524,288 & 4 & 36h \\
\midrule
Ours & \textbf{0.0008} & \textbf{0.9311} & 3 $\times$ (42, 64, 64) & 516,096 & 1 & 8h \\
\bottomrule
\end{tabular}
\caption{Quantitative comparison of reconstruction quality between autoencoder models: VAE-GAN~\cite{Pinaya2022BrainIG}, VQ-GAN~\cite{Khader2022MedicalDD} and our model. VAE-GAN*~\cite{Pinaya2022BrainIG} and VQ-GAN*~\cite{Khader2022MedicalDD} impose the same constraint [0,1] on the output as our model. The number of latent dimensions has been approximately matched across models to ensure fairness. }
\label{tab1}
\end{table*}
\begin{table*}[h]
\centering
\begin{tabular}{llcccccc}
\toprule
Method & Labels & Precision $\uparrow$ & Recall $\uparrow$ & Specificity $\uparrow$ & Accuracy $\uparrow$ & F1 Score $\uparrow$ \\
\midrule
\multirow{3}{*}{VQ-GAN} & ED & 0.8140$\pm$2e-3 & 0.7498$\pm$3e-2 & 0.9950$\pm$1e-3 & 0.9880$\pm$1e-3 & 0.7815$\pm$1e-2   \\
 & NCR/NET & 0.6545$\pm$3e-2 & 0.6849$\pm$3e-2 & 0.9972$\pm$2e-3 & 0.9948$\pm$1e-3 & 0.6682$\pm$2e-2 \\
& ET & 0.7127$\pm$3e-2 & 0.7557$\pm$2e-2 & 0.9968$\pm$1e-3 & 0.9944$\pm$1e-3 & 0.7333$\pm$2e-2 \\
\midrule
\multirow{3}{*}{Ours} & ED & {\bf0.9430}$\pm$3e-3 & {\bf0.9540}$\pm$2e-3 & {\bf0.9983}$\pm$2e-5 & {\bf0.9970}$\pm$1e-4 & {\bf0.9500}$\pm$2e-3   \\
 & NCR/NET & {\bf0.9250}$\pm$1e-2 & {\bf0.9321}$\pm$1e-2 & {\bf0.9993}$\pm$1e-4 & {\bf0.9988}$\pm$1e-4 & {\bf0.9273}$\pm$1e-2 \\
& ET & {\bf0.9183}$\pm$1e-2 & {\bf0.9455}$\pm$5e-3 & {\bf0.9992}$\pm$1e-4 & {\bf0.9987}$\pm$1e-4 & {\bf0.9350}$\pm$2e-3 \\
\bottomrule
\end{tabular}
\caption{Downstream segmentation task: quantitative performance comparison in 3D medical segmentation using the BraTS dataset, compared to the VQ-GAN model~\cite{Khader2022MedicalDD}. This table presents Precision, Recall, Specificity, Accuracy, and F1 Score for each segmented label, based on four repeated experiments. Training the VAE-GAN model~\cite{Pinaya2022BrainIG} on this dataset encounters out-of-memory issues.}
\label{tab2}
\end{table*}
\begin{figure}[h]
\centering
\includegraphics[width=0.95\columnwidth]{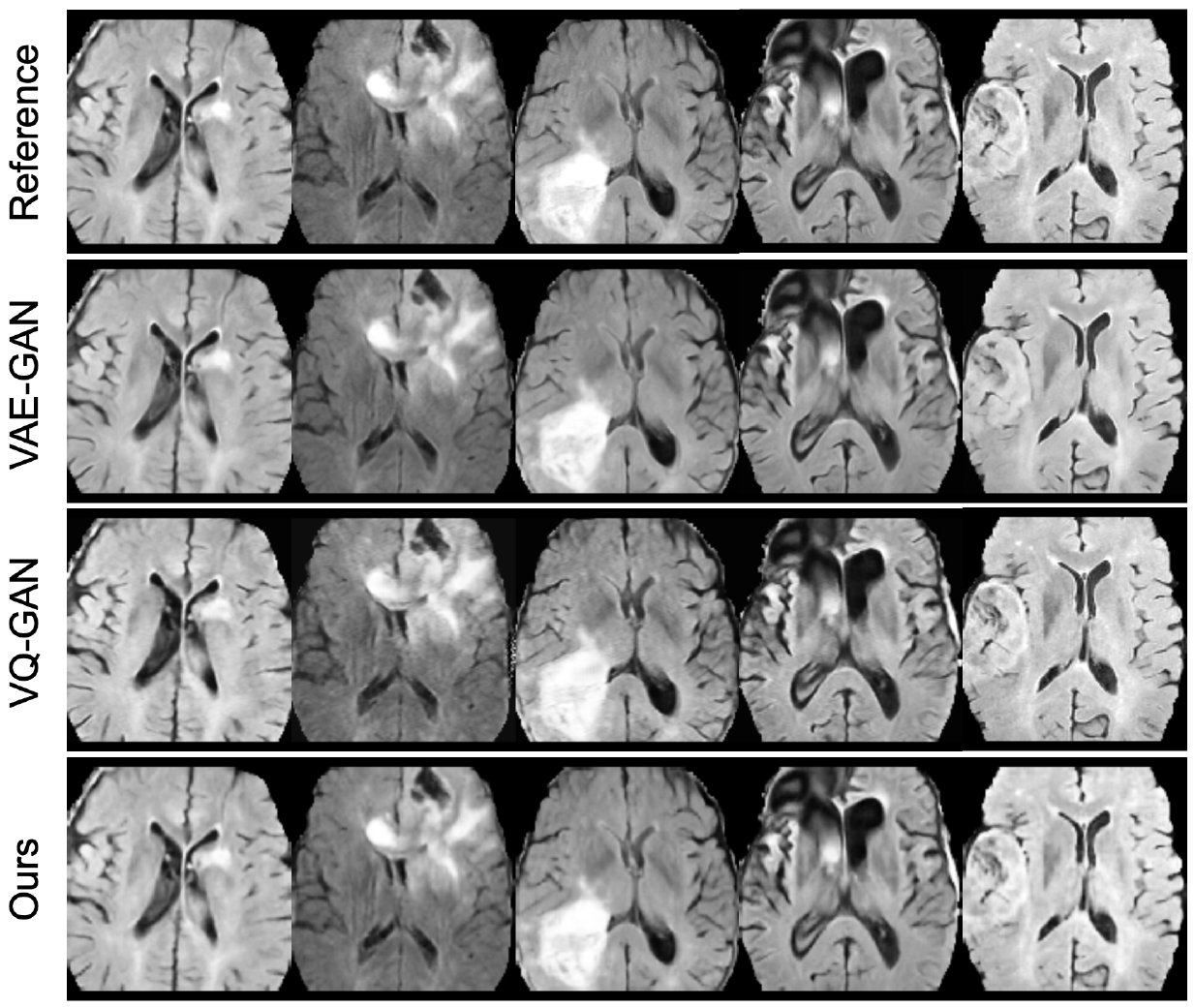}
\caption{Qualitative comparison of reconstruction quality between autoencoder models: VAE-GAN~\cite{Pinaya2022BrainIG}, VQ-GAN~\cite{Khader2022MedicalDD} and our model.}
\label{fig3}
\end{figure}
We compare our decoder-only autoencoder model's performance with VAE-GAN~\cite{Pinaya2022BrainIG} and VQ-GAN~\cite{Khader2022MedicalDD}. Figure~\ref{fig3} shows our model achieves remarkable reconstruction results, closely approximating the ground truth. Table~\ref{tab1} indicates lower MSE and 3D SSIM values, with faster training and less computational resource use, highlighting our model's efficiency and effectiveness in preserving image details.
\begin{figure}[h]
\centering
\includegraphics[width=0.95\columnwidth]{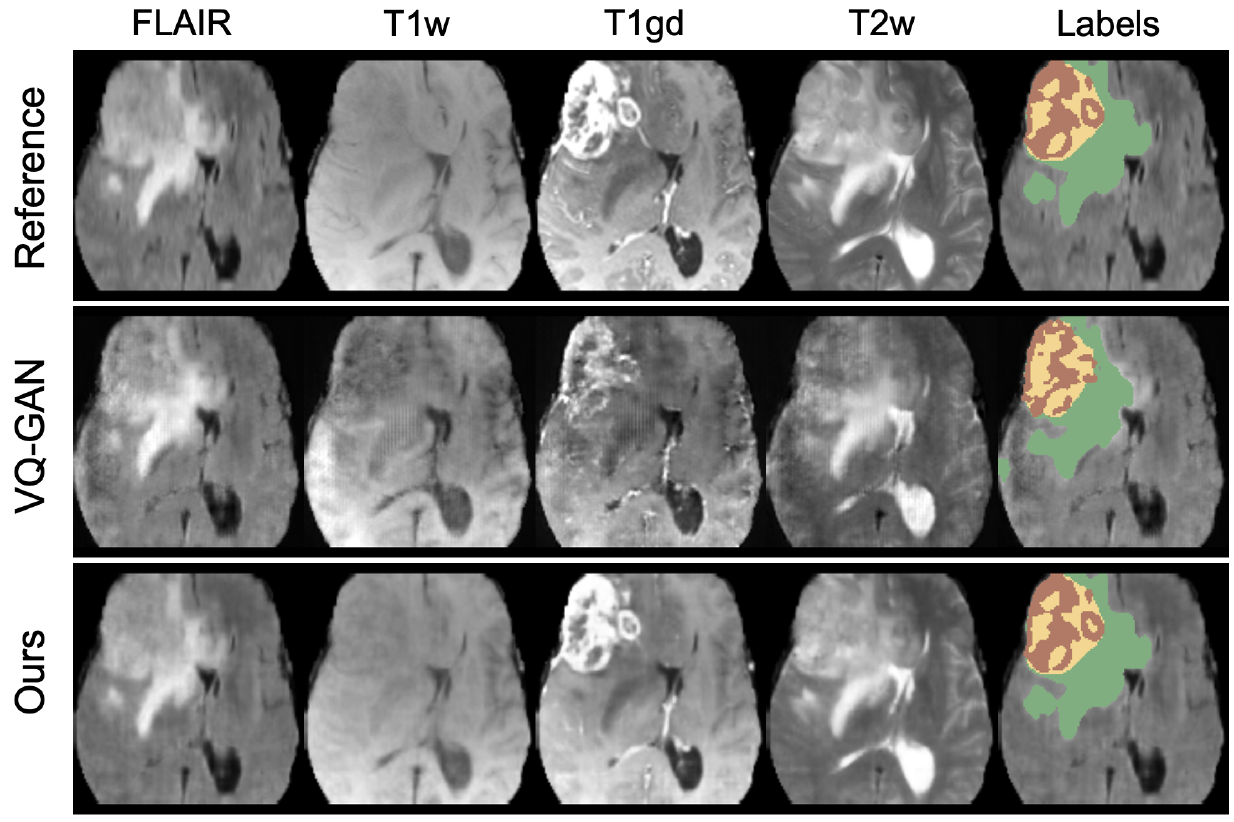}
\caption{Downstream segmentation task: evaluation of reconstructed image performance in multi-modal 3D segmentation, comparing our model to the VQ-GAN model~\cite{Khader2022MedicalDD}. ET label is the brown area, the ED label is the green area, and NCR/NET label is the yellow area~\cite{menze2014multimodal,bakas2018identifying}.}
\label{fig4}
\end{figure}
We generate multi-modal data using the same triplane representations to ensure our decoder-only autoencoder retains essential medical information. Due to out-of-memory errors, VAE-GAN~\cite{Pinaya2022BrainIG} couldn't be trained on the 4-channel BraTS dataset~\cite{menze2014,Simpson2019ALA}, so it was excluded from segmentation comparisons. Figure~\ref{fig4} and Table~\ref{tab2} show that our model preserves critical information for medical segmentation and scales more efficiently from single-channel to multi-channel configurations compared to VQ-GAN~\cite{Khader2022MedicalDD} while retaining crucial information.

\subsection{Generative Models}
\begin{figure}[h]
\centering
\includegraphics[width=1.0\columnwidth]{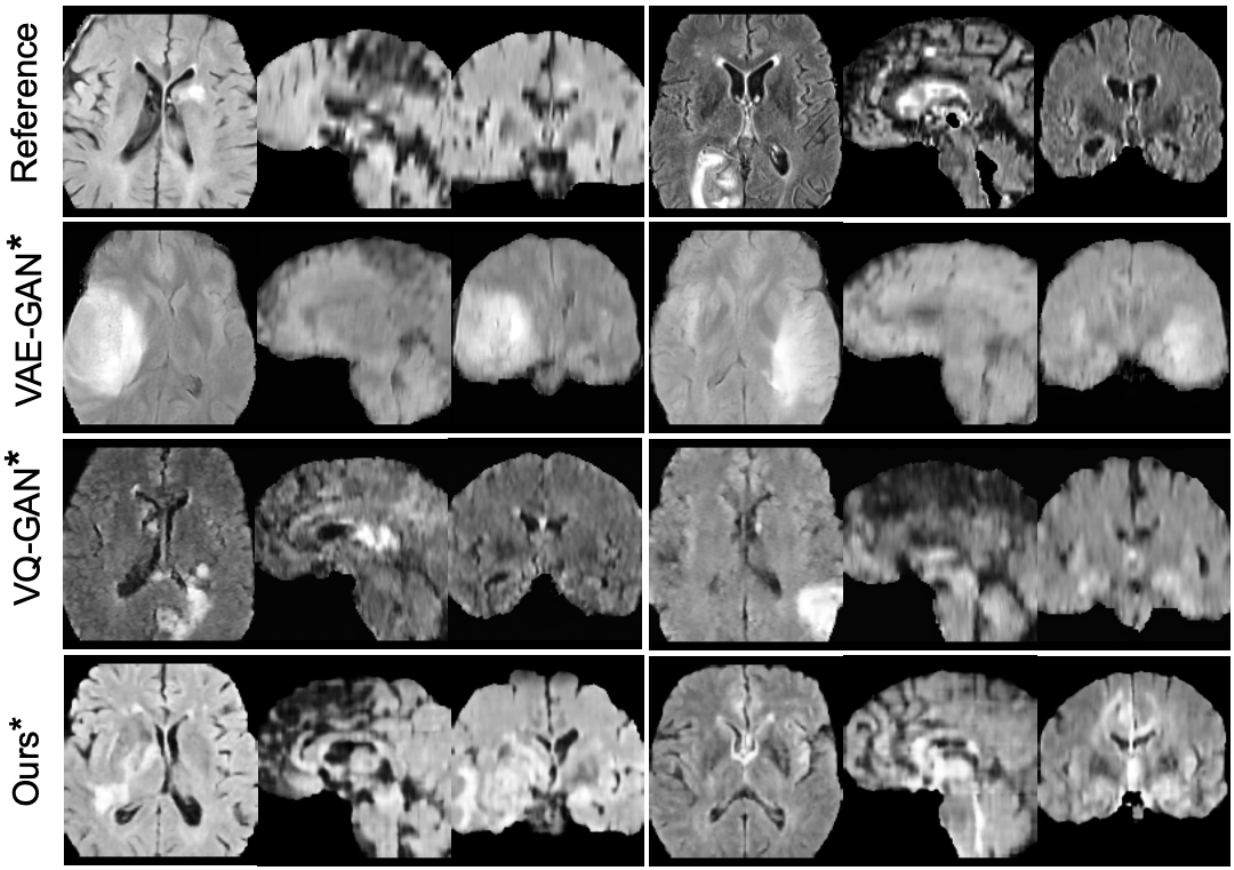}
\caption{Qualitative comparison of generative quality between VAE-GAN-LDM (VAE-GAN*)~\cite{Pinaya2022BrainIG}, VQ-GAN-LDM (VQ-GAN*)~\cite{Khader2022MedicalDD}, and our triplane-aware diffusion model in BraTS dataset.}
\label{fig5}
\end{figure}
\begin{table}[h]
\centering
\begin{tabular}{lc}
\toprule
 Method & Wasserstein distance $\downarrow$ \\
 \midrule
 VAE-GAN-LDM & 120$\pm$20 \\
 Ours & \textbf{36}$\pm$4 \\
 \midrule
 \midrule
 VQ-GAN-LDM & 160$\pm$20 \\
 Ours & \textbf{16}$\pm$5 \\
 \bottomrule
\end{tabular}
\caption{Quantitative comparison of generative quality measured by Wasserstein distance using a W-GAN critic~\cite{Arjovsky2017WassersteinG} for two baseline models and our generative model. Experiments were conducted three times using different random seeds to ensure reliability.}
\label{tab3}
\end{table}
We evaluate the performance of the generative model, between the VAE-GAN-LDM~\cite{Pinaya2022BrainIG}, the VQ-GAN-LDM~\cite{Khader2022MedicalDD} and our triplane-aware cross-attention diffusion model. Figure~\ref{fig5} demonstrates that the volumes generated by our model possess more detail and are closer to real images, compared to the results from the VAE-GAN-LDM~\cite{Pinaya2022BrainIG} and VQ-GAN-LDM~\cite{Khader2022MedicalDD}. From Table~\ref{tab3}, our model achieved a lower Wasserstein distance in the experiments, compared to these baseline models. This demonstrates that the generated data distribution from our model is closer to the real data distribution, surpassing these two baseline models' ability to create realistic samples. See more results in the appendix.

\subsection{Performance in High-Resolution Data}
\begin{figure}[h]
\centering
\includegraphics[width=1.0\columnwidth]{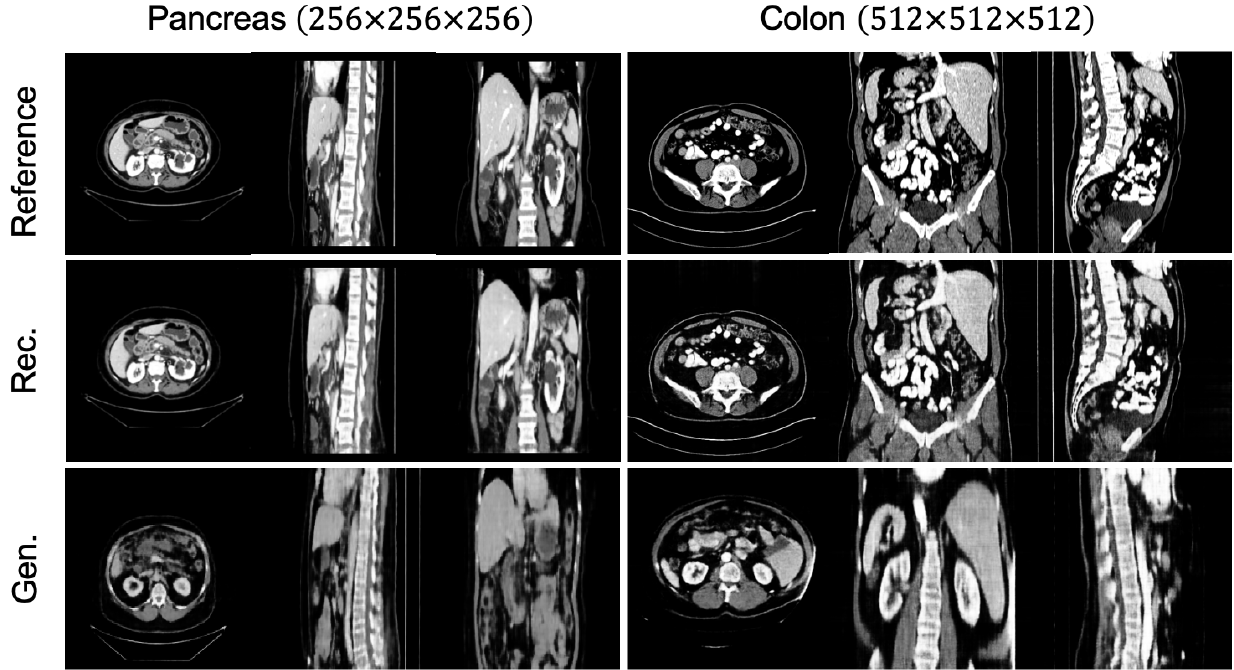}
\caption{Reconstruction quality and generative quality of our model in pancreas dataset~\cite{Simpson2019ALA} with $256\times256\times256$ resolution, and colon dataset~\cite{Simpson2019ALA} with $512\times512\times512$ resolution. $Rec.$ means the reconstruction output. $Gen.$ means the generative output.}
\label{fig6}
\end{figure}
When handling the high-resolution dataset, VAE-GAN-LDM\cite{Pinaya2022BrainIG} and VQ-GAN-LDM~\cite{Khader2022MedicalDD} suffer the out-of-memory problem due to significant increases in the number of parameters, more gradient storage, etc. In contrast, our model uses the decoder-only autoencoder model, which can significantly reduce the memory usage during training and inference to allow handling higher-resolution datasets than VAE-GAN~\cite{Pinaya2022BrainIG} or VQ-GAN~\cite{Khader2022MedicalDD}. Figure~\ref{fig6} present the reconstruction and generation quality of our model when handling high-resolution datasets. See more results in the appendix.

\subsection{Ablation Experiments}
\paragraph{Autoencoder Ablation.} We decreased the channel size for the triplane representation to utilize fewer parameters in the latent dimension to achieve similar results compared to the VAE-GAN diffusion model~\cite{Pinaya2022BrainIG} in terms of the performance for the autoencoder model. From Table~\ref{tab4}, our model achieves comparable reconstruction quality to the VAE-GAN~\cite{Pinaya2022BrainIG} while utilizing only its $48\%$ of the latent dimension parameters.
\begin{table}[h]
\centering
\begin{tabular}{llccc}
\toprule
Method & Params & Ratio & MSE $\downarrow$ & 3D SSIM $\uparrow$ \\
\midrule
 VAE-GAN* & 524,288 & 1.0 & {\bf0.0014} & {\bf0.8818}  \\
 \cmidrule(r){2-5}
 \multirow{4}{*}{Ours} & 221,184 & 0.41 & 0.0015 &  0.8914 \\
 & 245,760 & {\bf0.48} & {\bf0.0014} & {\bf0.8985} \\
 & 294,912 & 0.56 & 0.0012 & 0.9117 \\
 \cmidrule(r){2-5}
 & 516,096 & 0.98 & 0.0008 & 0.9311 \\
\bottomrule
\end{tabular}
\caption{Performance changes with decreasing channel size. Ratio refers to the latent dimensions of our models compared to the VAE-GAN model~\cite{Pinaya2022BrainIG}.}
\label{tab4}
\end{table}

\begin{figure}[h]
\centering
\includegraphics[width=1.0\columnwidth]{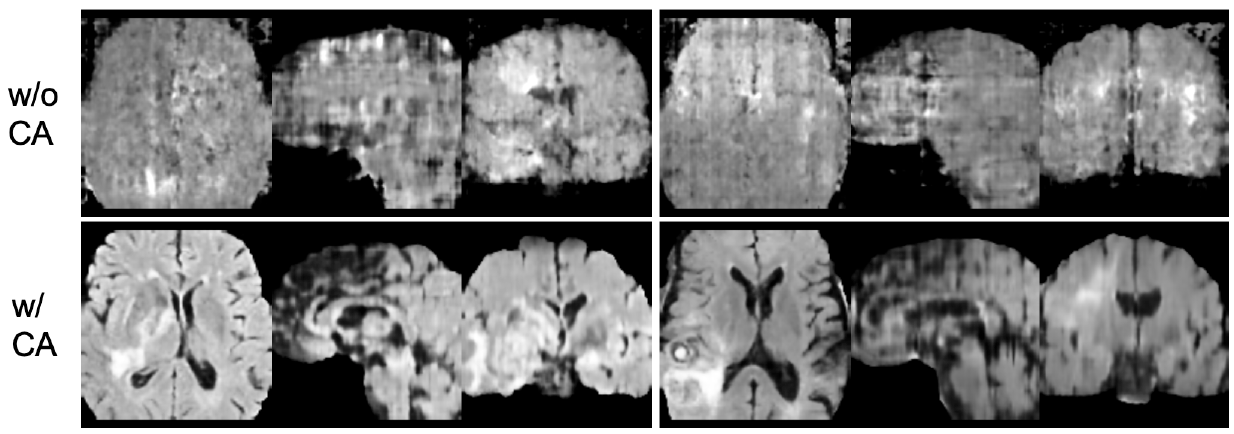}
\caption{Performance between the diffusion model with (second row) and without (first row) cross-attention layers.}
\label{fig7}
\end{figure}

\paragraph{Diffusion Ablation.} In Figure~\ref{fig7}, we conducted a comparison of diffusion models with and without the integration of cross-attention, revealing that cross-attention significantly improves the extraction of information from 3D volumes. Additionally, it enhances the consistency of the sampled triplanes, thereby contributing to more reliable and accurate reconstructions in our models.

\section{Conclusion}
In this paper, we introduce a novel and effective diffusion model called TCAM-Diff, triplane-aware cross-attention mechanisms for generating 3D medical datasets. We demonstrate how TCAM-Diff surpasses existing encoder-decoder methods by delivering superior reconstruction and generation quality with similar-sized latent spaces. Its decoder-only design enables efficient handling of high-resolution datasets without encountering memory issues. Extensive qualitative and quantitative experiments validate the feasibility and effectiveness of TCAM-Diff in generating high-quality 3D medical data. 

\section{Acknowledgments}
This research was supported by The University of Melbourne’s Research Computing Services and the Petascale Campus Initiative.

\bibliography{aaai25}

\end{document}